\documentclass[acmlarge, nonacm]{acmart}
\usepackage{setspace}
\usepackage{indentfirst}
\usepackage{wasysym}
\newrobustcmd{\emoji}[1]{\includegraphics[height=1.10\fontcharht\font`\A]{emoji_images/#1.jpeg}}

\title[ElectionRumors2022]{ElectionRumors2022: A Dataset of Election Rumors on Twitter During the 2022 US Midterms} 

\author{Joseph S. Schafer}
\authornote{Both authors contributed equally to this research.}
\email{schaferj@uw.edu}

\author{Kayla Duskin}
\authornotemark[1]
\email{kduskin@uw.edu}
\affiliation{%
  \institution{University of Washington}
  \country{USA}
}

\author{Stephen Prochaska}
\affiliation{%
  \institution{University of Washington}
   \country{USA}
}

\author{Morgan Wack}
\affiliation{%
  \institution{Clemson University}
   \country{USA}
}

\author{Anna Beers}
\affiliation{%
 \institution{University of North Carolina}
  \country{USA}
}

\author{Lia Bozarth}
\affiliation{%
  \institution{University of Washington}
   \country{USA}
}

\author{Taylor Agajanian}
\affiliation{%
  \institution{University of Washington}
   \country{USA}
}

\author{Mike Caulfield}
\affiliation{%
  \institution{University of Washington}
  \country{USA}
}

\author{Emma S. Spiro}
\affiliation{
  \institution{University of Washington}
  \country{USA}
  }

\author{Kate Starbird}
\affiliation{%
  \institution{University of Washington}
  \country{USA}
}

\usepackage{longtable}
\usepackage{tabu}
\renewcommand{\shortauthors}{Schafer \& Duskin et al.}

\begin{document}

\begin{abstract}
  \noindent Understanding the spread of online rumors is a pressing societal challenge and an active area of research across domains. In the context of the 2022 U.S. midterm elections, one influential social media platform for sharing information --- including rumors that may be false, misleading, or unsubstantiated --- was Twitter (now renamed X). To increase understanding of the dynamics of online rumors about elections, we present and analyze a dataset of 1.81 million Twitter posts corresponding to 135 distinct rumors which spread online during the midterm election season (September 5 to December 1, 2022). We describe how this data was collected, compiled, and supplemented, and provide a series of exploratory analyses along with comparisons to a previously-published dataset on 2020 election rumors. We also conduct a mixed-methods analysis of three distinct rumors about the election in Arizona, a particularly prominent focus of 2022 election rumoring. Finally, we provide a set of potential future directions for how this dataset could be used to facilitate future research into online rumors, misinformation, and disinformation.
\end{abstract}

\maketitle

\section{Introduction}
Online rumors, and related phenomena such as misinformation and disinformation, have become increasingly important to understand and address, with prior work framing misinformation and disinformation as urgent problems \citep{Calo2021-ca} which need an interdisciplinary and integrated “crisis discipline” approach to research \citep{Bak-Coleman2021-xj}. Rumors are often a byproduct of collective sensemaking processes, whereby people come together, frequently in online spaces, to make sense of ambiguous and/or uncertain information \citep{shibutani_improvised_1966,arif_how_2016}. Events such as natural hazards, public health crises, and elections are frequent catalysts for rumoring behavior, particularly given that they frequently draw global attention and involve high levels of informational uncertainty. Unsurprisingly, a growing body of research has focused on understanding sensemaking, rumoring, and misinformation in the context of democratic elections, both within the United States (U.S.) \citep{Bovet2019-qx, Jones-Jang2021-ml,Kennedy2022-qv,Oehmichen2019-dd} and globally (e.g. \citep{Akbar2022-un, Mendoza2023-eb, Recuero2020-mb}).

In the U.S. context, research on election-related misinformation and rumors has often focused on presidential elections, such as in 2016 and 2020 \citep{Bovet2019-qx,Kennedy2022-qv,Oehmichen2019-dd,Guess2019-oj, Grinberg2019-wy, Sharma2022-zx, Moore2023-wy}. However, midterm elections are also sites of potential election rumoring, and are of critical importance to governmental composition, despite typically commanding somewhat less public attention and participation. The 2022 U.S. midterm elections consisted of nationwide elections for all members of the U.S. House of Representatives and a partial set of members of the U.S. Senate, as well as for many state and local races (e.g. the governor's race in Arizona). Documenting election process-related rumors in this specific context allows for important gains to research understanding; not only can we better understand the dynamics of rumors online by expanding the cases studies within the research community, but we can also consider whether phenomena observed in presidential elections apply to U.S. non-presidential election contexts as well.

In this paper, we aggregate and describe a set of 135 rumors that spread on Twitter (now renamed X)\footnote{For clarity we will use terminology as it existed at the time of data collection, e.g. Twitter, tweets, retweets, quote tweets, and reply tweets} --- at the time, a notable social media platform for real-time information and news sharing in the U.S. --- during the 2022 U.S. midterm elections. We curate a set of posts that include original tweets, retweets, quote tweets, and reply tweets that correspond to each rumor, which in aggregate comprises 1.81 million posts. We additionally include information on the web domains that were cited in content pertaining to each rumor, and the geographical location (U.S. state or states) that is the focus of each rumor. We describe the process used for creating and curating a comprehensive (high recall) and low-noise (high precision) sample of tweets for each rumor. We present several empirical analyses of this data, first through five preliminary descriptive statistical analyses, positioned in comparison with a similar dataset on the 2020 U.S. elections, previously published in \citep{Kennedy2022-qv}. Next, to demonstrate the utility of these data for more in-depth research, we feature a mixed-methods, case study analysis of three rumors focused on the state of Arizona. We conclude by outlining future work that could build off of this dataset and accompanying analyses, along with reflections on the ethical considerations and methodological limitations of using and researching with these data. 

Importantly, given recent changes in data accessibility, it is increasingly difficult to study social media though large-scale observations and data analysis; it is now nearly impossible to estimate changes in online communication patterns over time \citep{Weatherbed2023-rr,Davidson2023-wd}. The time period of this study captured a high-impact public event while the Twitter platform still had comparatively open data accessibility, which makes observing and measuring rumors in this context even more compelling. As an important hub for discussion during the 2022 elections, Twitter provides one window into online rumoring, complementing studies of other platforms. This study, when paired with prior work, provides rare opportunity for insight into how online election-related discourse has changed over time, and offers implications for future election cycles.

\section{Background}
Throughout this paper, we use the framing of election-related \textit{rumors}. In this section, we briefly define this term and discuss how it relates to other forms of information --- namely \textit{misinformation} and \textit{disinformation}. We then provide background regarding the study of rumors online and in the context of election administration. 

\subsection{Terminology and Definitions}
Rumors are stories that contain information that is unverified or incomplete at the time of dissemination. They are a key component of conveying important information through a population \citep{pendleton1998rumor}. From a sociological perspective, rumoring is a crucial social process --- a form of collective problem-solving that has taken place consistently throughout history \citep{shibutani_improvised_1966}. Shibutani explains that rumors involve the pooling of intellectual resources to provide meaningful interpretation to an otherwise ambiguous situation \citep{shibutani_improvised_1966}. Situations characterized by high uncertainty and elevated anxiety are the prime environment for rumors to emerge \citep{anthony1973anxiety,rosnow1980psychology}. As situational uncertainty resolves, the information shared through rumors, while initially unverified, may later be shown to be true or false, or as time goes on may remain ambiguous --- being neither substantiated nor disproven. 

Rumors that turn out to be false or misleading can be considered \textit{misinformation}. Misinformation is inaccurate information that may be shared without the intent to mislead or harm audiences, or potentially without awareness that the message contains falsehoods \citep{Jack2017-bg, Freelon2020-kz,Calo2021-ca}. Closely related is \textit{disinformation}, denoting false or misleading information that is shared deliberately to further a particular goal such as monetary profit or political gain \citep{Jack2017-bg, Freelon2020-kz}. A disinformation campaign may consist of strategic amplification of a mixture of true, false, and misleading content that contributes to a desired narrative. Disinformation campaigns have been known to amplify organic rumors, as well as seed new rumors, as art of their aims. In these cases, disinformation may be spread unwittingly by audiences unaware of its deceptive nature \citep{Rid2020-oi}. 

The dataset described in this paper is comprised of a wide range of initially unverified content shared on Twitter --- some posts contain early reports of real events related to the election, other posts include soundly disproven claims, while some posts add misleading framing to factual events. Given this range, we find the term rumor to be the most accurate and useful descriptor for the collection. Conceptualizing these election-related narratives as rumors serves to highlight the sensemaking aspect of online communication, even when the topic of discussion ultimately turns out to be lacking veracity. This acknowledgment of uncertainty is key to understanding and productively discussing modern information systems \citep{Spiro2023-kv}. 

While we use rumors as the conceptual framework for this work, we acknowledge connections to research in mis- and disinformation, particularly given that this work specifically focuses on rumors with a false, misleading, or unsubstantiated element. Given the fluctuating understandings of terms, previous work exploring false and misleading online narratives use different terms for the mixed-veracity messages present in online social discourse. For example, Kennedy et al. uses `misinformation' as an umbrella term --- inclusive of rumors, misinformation, and disinformation --- in their study of stories shared online during the U.S. 2020 election \citep{Kennedy2022-qv}. Similarly, in their work characterizing user engagement with conspiratorial topics on Twitter during the 2020 U.S. election, Sharma et al. use `disinformation' as term to capture what they call `distorted narratives' which include unreliable or conspiratorial claims regardless of the intent of the content's author \citep{Sharma2022-zx}. In this work we do not attempt to distinguish individual pieces of content as true or false (e.g. to label posts as misinformation) nor as intentionally or unintentionally misleading (e.g. to label posts as disinformation). Rather, we organize content into distinct rumors, acknowledging the overlapping functions and understandings of rumor, stories, misinformation, and disinformation. 

\subsection{Online Rumoring}
Online social media platforms have changed how people consume news, learn about crises, and keep up with current events. In the digital age, rumors have the potential to spread faster than ever and reach broader audiences \citep{doerr2012rumors,Sunstein2009-vi}. At the same time, analyzing the digital trace data available in these systems has allowed researchers to gain insight into how rumors spread through networked sociotechnical systems and populations. Research in this space has focused heavily within the domain of crisis informatics -- the study of how citizens respond to and make sense of emerging crises through online communication \citep{palen2007crisis}. Studies have helped illuminate online rumor propagation \citep{arif_how_2016, Zeng2016-hj} and correction \citep{arif2017closer, starbird_journalists}, along with their role in collective sensemaking \citep{Starbird2016-ty} in the context of diverse public crises ranging from natural hazards to acts of violence such as shootings or hostage situations. 

The affordances of online social media platforms allow users to both share and seek out emerging information extremely rapidly, and without the gatekeeping mechanisms of traditional media communication (though these networks undoubtedly have their own access constraints \citep{keegan_egalitarians_2010}). In fact, \citep{doerr2012rumors} show that the very structure of online social networks (specifically Twitter) facilitates faster rumor propagation than other networks of similar size and density. Additionally, the novelty and salience of information contained in rumors may contribute to their propensity to spread rapidly online. In a large-scale study of online news, \citep{Vosoughi2018-pi} show that false news travels farther, faster, and more broadly on Twitter than true news and is also more novel and surprising to audiences than true news. As online populations collectively seek to make sense of an uncertain, stressful, or novel situation, rumors emerge to provide explanation and fill information voids -- even if their validity is unknown. 

\subsection{Election Administration Rumors}

One context ripe for rumors is election administration, given the high degree of uncertainty and consequence associated with them. As such, rumors during and about election administration have gained increased attention in the last decade in particular. In the U.S., there is a strong political element to the formation of collective identity -- one which relies on deep stories that resonate with their intended audience, regardless of factuality \citep{polletta_deep_2017}. Deep stories of voter fraud and election malfeasance have become increasingly prominent in recent years \citep{prochaska_mobilizing_2023}. In recent work, researchers also found that in states with more restrictive ballot counting laws, and therefore longer periods of uncertainty, rumors and misinformation were more prevalent during the 2020 U.S. election \citep{wack_working_2023}. This confluence of factors indicate that rumoring about the administration of elections is not likely to be a temporary trend, but rather an expected element of information ecosystems that we seek to better describe and conceptualize. 

Other researchers have also focused on studying social media discussions of elections and election administration by compiling datasets of social media posts. Notable related datasets include the \textit{VoterFraud2020} dataset focused on \#VoterFraud and related hashtags during the 2020 election \citep{Abilov2021-ki}, the \textit{ElectionMisinfo2020} dataset focused on 2020 election misinformation using mixed-methods curation described in \citep{Kennedy2022-qv}, the \textit{\#Election2020} dataset focused on broad coverage of 2020 election discussions on Twitter in \citep{chen_election2020_2022}, and the \textit{MEIU22} dataset of posts broadly related to the 2022 U.S. midterm elections across multiple platforms in \citep{Aiyappa2023-rp}. Our dataset adds to these resources by using a similar narrowly-scoped, low-noise dataset akin to \textit{ElectionMisinfo2020} \citep{Kennedy2022-qv}, while being focused on the particular time that \textit{MEIU22} \citep{Aiyappa2023-rp} studies, filling an important gap that existing datasets do not cover. In addition to its utility for understanding discussions of election administration rumors during the 2022 midterms on Twitter in isolation, our dataset is useful for comparative studies with either of those datasets --- for example, to understand how rumor-based conversations differed from more general discourses as covered in \textit{MEIU22} \citep{Aiyappa2023-rp}, or how rumor-based online engagement evolved from 2020 to 2022 by comparing to \textit{ElectionMisinfo2020} \citep{Kennedy2022-qv}. Studying Twitter during this time period is particularly important, as it was in the beginning stages of significant business and ownership changes \citep{conger_elon_2022}. In our analysis section, we will demonstrate several of these comparative analyses to \textit{ElectionMisinfo2020} \citep{Kennedy2022-qv}. 

\section{Tweet Collection \& Rumor Identification}
\begin{figure*}
\centering
\includegraphics[width=.95\linewidth]{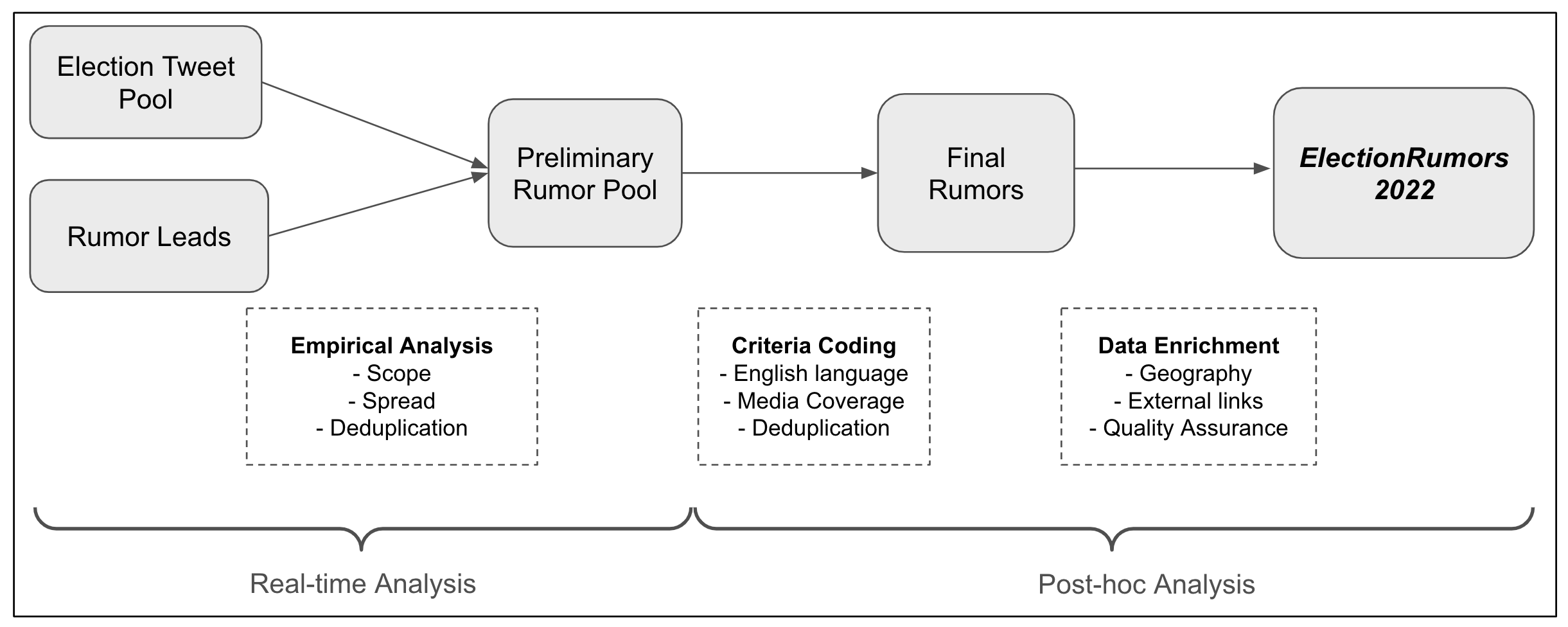}
\caption{A diagram showing the process of curating this dataset}
\label{fig:process}
\end{figure*}

In this section we detail the collection of tweets broadly related to the midterm election, how we identified emergent election rumors in real time and matched tweets to their corresponding rumor, and finally how we evaluated quality at both the rumor and tweet level. Figure~\ref{fig:process} depicts this iterative procedure wherein we employ empirical and qualitative evaluation to a broad dataset to create a rich dataset that enables assessment of rumoring about the administration of the 2022 U.S. election. 

\subsection{Collection of Election Tweets}
Throughout the months leading up to and following the 2022 U.S. midterm elections, our team collected a broad set of election-related data using Twitter's V1.1 streaming API\footnote{\parbox [t] {\linewidth} {\url{ https://web.archive.org/web/20220307124146/https://develope r.twitter.com/en/docs/twitter-api/v1}}}. We did so using a list of keywords, phrases, and hashtags related to the election (e.g. \textit{ballot}, \textit{vote}) as well as to narratives common in election rumoring (e.g. \textit{fraud}, \textit{tabulator}). These terms were selected by a team of researchers with contextual expertise studying previous elections and informed by prior work conducted during the 2020 U.S. elections \citep{Kennedy2022-qv}. The keywords were designed to capture a comprehensive dataset of tweets related to voting, election materials, procedures, results, and claims of election fraud. The full list of keywords used in collecting the general election tweets is included in the Appendix in Table \ref{tab:keywords}. Modifications (additions and deletions of keywords) were made to this list to adapt to emerging trends and narratives, and these modifications are also noted in the Appendix in Table \ref{tab:keywords}. 

To reduce the impact of rate-limiting, keywords were divided among nine separate collectors; each collector had its own streaming credential attached to different members of our research team (Twitter's V1.1 API allowed up to max 50 tweets per second collection for any one credential). It is important to note that the collectors were intermittently rate-limited in the two weeks prior to the 2022 Midterm election, and were consistently rate-limited on Nov 7th and Nov 8th. Additionally, one of the collectors went briefly down on Nov 2nd due to a credential issue that was quickly resolved. 

We approximate the comprehensiveness of the resulting dataset of collected election-related tweets using missing retweets. That is, let tweet $i \in {SharedOrigTweets}$ where $SharedOrigTweets$ contains all the original tweets that have at least 1 retweet in the dataset. We denote $|i|_{obs}$ as the number of observed retweets of $i$ in the dataset. Further, let $|i|_{exp}$ be the expected number of retweets for tweet $i$ (i.e. the number of retweets that should be present in the dataset). We derive $|i|_{exp}$ by taking the earliest and the latest versions (i.e., timestamped) of $i$ observed, and compute the difference in the recorded retweet counts (embedded in the meta-data) of the two versions. Finally, the fraction of missing retweets of $i$ is defined as the difference between $|i|_{exp}$ and $|i|_{obs}$. We find that 12\% of retweets were missing from the dataset overall; 20\% of retweets were missing during the Midterm election week (11/02/2022 to 11/08/2022). This is consistent with prior literature which demonstrated that the streaming API can miss significant portions of data \citep{pfeffer_this_2023, wang_should_2015}. However, missing data from the Streaming API should not significantly impact topic coverage \citep{pfeffer_this_2023}.

To improve data comprehensiveness, for original tweets that have at least one retweet captured by our collectors but are themselves missing from the dataset, we used a separate process to attempt to backfill these original tweets. We backfilled approximately 2.38 million tweets (of which approximately 1,450 were included in the final, curated rumor dataset). This collection process resulted in an initial election tweet set of 446 million tweets.

\subsection{Identifying Rumor Leads}
During a period of ten weeks between 9/19/2022 and 12/01/2022, a team of 15 undergraduate research assistants worked in daily shifts to actively observe online discourse about the election and document instances of online content related to U.S. election integrity. Similar to the collection of the initial election-related tweet dataset, initial rumor leads were focused on enabling rapid, broad coverage of online discourse which could involve rumors related to the election (at this stage, the team emphasized coverage rather than precision, as we would curate this set further at a later step). Researchers first conducted advanced search queries on Twitter using keyword queries and made note of any content with the potential to be an election rumor. Some baseline keyword queries were used daily throughout the data collection period (e.g. `(ballot OR ballots)'), while others were customized to exogenous events such as news articles or current events related to the election. Content containing potential rumors was logged if it met two scoping criteria: 1) the potential rumor contained a new (or newly prominent) claim that hadn't been logged by the team previously and 2) the claim met one or more of the topical scope criteria. Though the initial scope of our real-time analysis included other categories, such as threats to election personnel, for this dataset, we focus on rumors which either are likely to deprive someone of their vote (which would include procedural interference, participation interference, or solicitation of fraud), or to cast doubt on the integrity of the election processes and/or the accuracy of election results. Once logged, the undergraduate assistants conducted initial manual analysis to identify any related content on Twitter (or other platforms)\footnote{Searches of these other platforms, including Facebook, Instagram, TikTok, and Telegram, were used to inform related rapid-response work, but was not used for the Twitter-focused dataset described in this paper.} and summarize the claims present.

\subsection{Constructing a Preliminary Rumor Pool}
During a period of eight weeks between 9/26/2022 and 11/22/2022 a group of 17 researchers, including nine of the authors, worked in team shifts refining and investigating the logged rumor leads. Teams used qualitative and quantitative analysis to either flesh out the leads into preliminary rumors, or discard them for not meeting scope --- the most common reasons for discarding a rumor were identification of a duplicate rumor, or having very low spread on Twitter. Qualitatively, teams searched for relevant news coverage, related social media content, and official sources or fact-checks when available. To scope each rumor, researchers used an iterative, manual process to develop search queries to identify tweets in the election tweet pool related to each rumor (this process is similar to that described in \citep{Kennedy2022-qv}). Researchers identified a date range and combined keywords, twitter IDs, and/or URLs connected to the rumor using boolean operators to construct queries that captured as much of the relevant discourse as possible while minimizing noise. This allowed for analyses of rumor virality, breadth of spread, and identification of key messengers of each rumor. Key findings surfaced from these analyses were shared with journalists and the public during this time period through blog posts\footnote{https://www.eipartnership.net/blog} and Twitter threads\footnote{\url{https://twitter.com/EI_Partnership}}. 


\subsection{Rumor Criteria Coding}
Following the team's real-time daily analysis efforts, our team conducted post-hoc qualitative coding on the set of rumors during January and February, 2023 to further ensure de-duplication and consistency in meeting specified inclusion criteria. This assessment of the rumors involved three members of the research team with deep contextual knowledge of election related rumors, and high familiarity with the dataset. The key criteria met by all rumors included in the dataset presented here are as follows: 
\begin{itemize}
    \itemsep0em 
    \item Online discussion about the rumor on Twitter is primarily in English. 
    \item The rumor pertains to the 2022 U.S. midterm election, rather than a prior election cycle or an election in another country. 
    \item The rumor is either unsubstantiated or contradicted by authoritative sources (discussed below).
    \item The rumor is likely to deprive a person of their vote (e.g. false information about where or how to vote) or likely to delegitimize election results. 
    \item The rumor has non-trivial circulation on Twitter (e.g. more than one original tweet).
    \item The majority of discourse online relevant to this rumor occurred within the date range 9/5/2023 - 12/1/2023.

\end{itemize}

Notably, the post-hoc rumor criteria coding included searching for relevant authoritative sources (e.g. fact-checks, general reporting, official government communication) pertaining to each rumor. The purpose of this was to classify discourse around claims as either significantly substantiated, unsubstantiated, misleading/false, non-falsifiable, or primarily fact-check/correction. This was important in revisiting claims that were uncorroborated at the time of initial analysis but that later were shown to be true or false. 
We only include rumors found to be false, misleading, or unsubstantiated in the final set included for publication.

\subsection{Tweet Quality Assurance}
We evaluate the quality of each set of tweets associated with each rumor. First, we used the queries developed during the real-time analysis of rumors to identify tweets related to each rumor. For each rumor, we consider two tweets samples: the ten most-retweeted tweets, and a random sample of size ten of all other tweets excluding retweets (i.e., original tweets, quote tweets, and reply tweets). The two first authors qualitatively coded each tweet in the samples as either related or unrelated to the rumor with which it was meant to be associated\footnote{Tweets consisting of corrections or fact-checks of a rumor were coded as relevant, though these types of content are rare in the dataset.}. We set an acceptance criteria that nine out of the ten most-retweeted tweets returned by the query, and at least eight out of a random sample of ten non-retweet tweets must be related to the rumor. Any queries that did not meet that threshold were iteratively updated, re-sampled, and re-coded until this criteria was met. We further adjusted the queries based on temporal volume plots (for an illustration of such plots for individual rumors, see Figures \ref{fig:az_case_1}, \ref{fig:az_case_3}, \ref{fig:az_case_2} later in our analysis), to make sure that we did not prematurely cut off a rumor before its spread had significantly diminished or died out, or begin it after significant discussion had already occurred. Queries were also refined to minimize overlaps between rumors by identifying tweets that appeared in multiple rumors and updating queries so that they only matched the relevant rumor, if only one should have been included. Some overlaps remain, as some tweets do reference multiple rumors and should therefore be associated with both. Most rumors exceeded the minimum threshold for tweet-level precision. Across all incidents and their respective final queries, 99\% of the ten most-retweeted tweets were coded as correctly associated with the rumor, and 96\% of the ten random tweets were found to be correctly associated. 
 
\section{Data}
To summarize, we collected tweets corresponding to a set of 135 rumors about the 2022 U.S. Midterm election from September 5th, 2022 to December 1st, 2022. The overall dataset contains approximately 1.81 million tweets (88.0\% of which are retweets), and approximately 427,600 unique users. We also include several data enrichments, detailed below.

\emph{\textbf{Geographic Location:}}
To gauge where rumors were most concentrated geographically, we identify the state(s) of interest for each rumor. To do so, each rumor was coded with up to three states that were the focus of discussion within the narrative. If there were more than three states of equal relevance, or if the rumor did not have any geographic components, the rumor is coded as "General". Three of the 135 rumors were coded as relevant to both a specific state and as "General", as they had significant components related to particular states as well as to broader narratives about the election.

\emph{\textbf{External Links:}} \label{link_process}
To understand what external sources were used in discussions of rumors we consider links to external sites using URLs present in each tweet; for each shortened link (e.g. https://t.co/xxxxx) we used the pycurl library\footnote{http://pycurl.io/docs/latest/index.html} to unravel it to obtain the complete URL. We also extract the domain name from the complete URL using a regular expression.

\emph{\textbf{Media Coverage:}}
For each rumor we sought out relevant media coverage or fact-checks identified during the criteria coding. The included links are not an exhaustive list, and are included in the dataset to provide additional context for each rumor when possible. 12 of the 135 rumors did not have directly-identified sources linked during criteria coding.

\subsection{Data Format}
\begin{figure*}
\centering
\includegraphics[width=\linewidth]{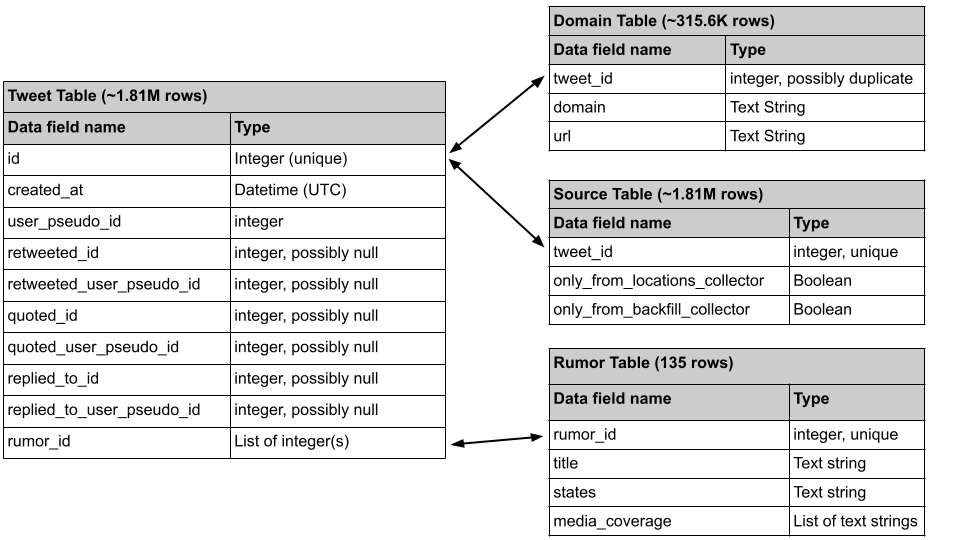}
\caption{A diagram of the data and feature relations contained in our dataset. Columns that are joinable across tables are linked via arrows.}
\label{fig:schema}
\end{figure*}

The dataset released in conjunction with this paper consists of three data types: \textit{rumors},  \textit{tweets}, and \textit{URLs}. For each \textit{rumor}, associated metadata include: an identifying number, the state(s) that the rumor was focused on, a short title summarizing what the rumor was claiming, and one or more URLs of relevant news coverage or fact-checks regarding that rumor. For each \textit{tweet}, metadata provided are: tweet ID (as provided by the Twitter API), a pseudonymized user ID, and the corresponding rumor ID number (3,640 tweets, (or 0.201\% of the dataset) referenced multiple rumors --- as a result, the rumor ID number column is stored as a list). We also release equivalent IDs and user pseudo-ids for referenced tweets (tweets which were retweeted, quoted, and/or replied to), as well as if a tweet was collected solely off of location-based keywords or backfilling, for researchers needing disaggregation of particular sources (for further discussion, see the appendix). For each \textit{URL}, the associated tweet ID is given. Figure \ref{fig:schema} summarizes the dataset features visually.

\subsection{Data Sharing}
We aim to make this dataset meet the FAIR principles \citep{wilkinson2016fair}, while also respecting both privacy concerns and the Twitter Terms of Service (at the time of collection). We make the dataset findable and accessible by hosting it in a Zenodo repository \citep{Schafer2024-dl}, allowing researchers to easily access and cite this resource. We also make it interoperable by releasing the standard tweet IDs for all posts contained within, so that other Twitter collections during this time (such as the Twitter component of \citep{Aiyappa2023-rp}) can be cross-referenced. We hope to catalyze data reuse by outlining several ideas for other projects in the discussion section, and by documenting how this data was collected herein so that researchers can evaluate the utility of this data for their own projects. Due to the reduced access of the Twitter API for rehydrating tweets, we will make subsets of hydrated data available to researchers upon reasonable request. However, both Twitter's Terms of Service and privacy concerns for those whose data we compile here prevent release of a fully hydrated dataset.

\section{Quantitative Analyses}
We perform five preliminary quantitative analyses, four of which include comparisons to a comparable dataset focused on the 2020 U.S. presidential election \citep{Kennedy2022-qv}. First, we analyze basic descriptive features of the data such as the temporal volume of tweets and largest rumors. We then conduct two analyses on the distribution of tweet topics --- by geographic focus and by user political partisanship. Next, we describe the prevalence of links to external domains, and differences in link-sharing behavior from the 2020 U.S. presidential election. Finally, we conduct an analysis of how heavily concentrated attention and activity were on the most-engaged-with and most-active accounts. 

\begin{figure}
\centering
\includegraphics[width=.95\linewidth]{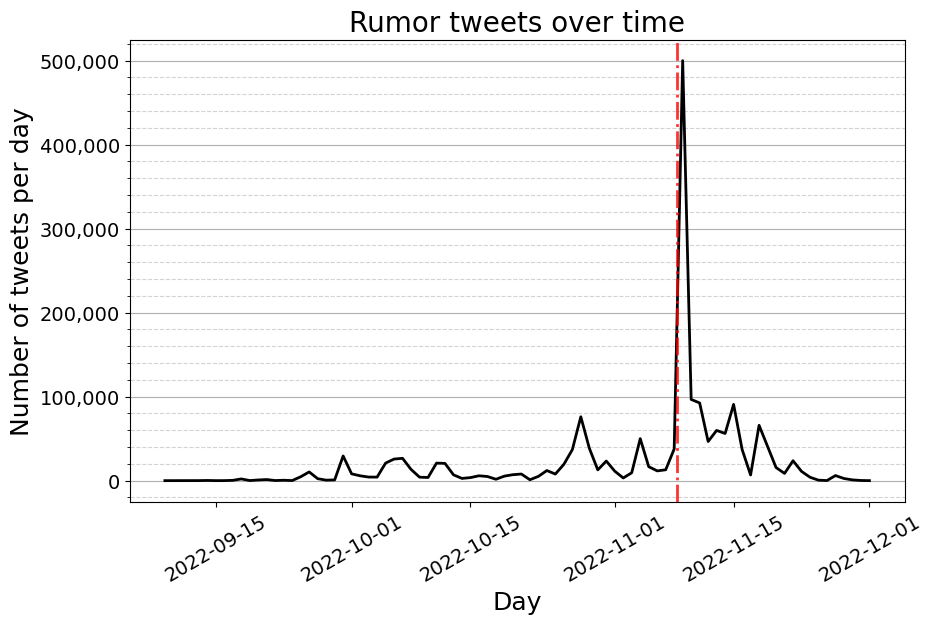}
\caption{A timeline plot showing the number of rumor tweets per day in the dataset. The start of Election Day (November 8, 2022, at 7:00 UTC) is denoted with a red vertical dashed line.}
\label{fig:timeline}
\end{figure}

\subsection{Descriptive Statistics}
First, we analyze the rate of tweets over time, with the daily count of rumor-related tweets presented in Figure \ref{fig:timeline}. The most striking feature is the pronounced spike in rumor circulation surrounding Election Day. Following Election Day, the rate of rumoring subsided quickly, with tweet volume dropping to less than a fifth of its Election Day spike in approximately 24 hours. By contrast, prior work in 2020, as shown in Figure 4 of \citep{Kennedy2022-qv}), demonstrated that election rumors were highly engaged with for multiple weeks after Election Day.

We describe the top set of rumors by tweet volume, which is documented in Table \ref{tab:top_rumors}. As this table illustrates, a significant number of the rumors which were most prevalent focused on the state of Arizona, prompting us to investigate how the geographic focus of rumor-related tweets were distributed.

\begin{table*}
    \renewcommand*\arraystretch{1.2}
    \centering
\begin{tabular}{|p{.4cm}|p{2.2cm}|p{2cm}|p{6cm}|p{2.5cm}|}
\hline
 & \# of Tweets & Rumor ID & Event of Rumor Focus & AZ Specific \\
\hline
1 & 224,384 & 66 & Maricopa machines not scanning ballots, Arizona & Yes\\
\hline
2 & 110,681 & 129 & Kari Lake shares stories of election day voters experiencing issues, Arizona & Yes\\
\hline
3 & 105,951 & 106 & Ballot counting speed comparisons between Arizona, Florida & Yes\\
\hline
4 & 62,576 & 9 & Konnech ties to China & No \\
\hline
5 & 58,505 & 126 & Republicans win downballot but not gubernatorial race, Arizona & Yes\\
\hline
6 & 54,905 & 64 & DOJ will observe election in 24 states & No\\ 
\hline
7 & 51,312 & 44 & Graphic showing election winner aired early, Arizona & Yes\\ 
\hline
8 & 49,315 & 6 & Louis DeJoy in charge of mail-in ballots & No\\ 
\hline
9 & 48,273 & 55 & Milwaukee elections official fired for election fraud, Wisconsin & No\\
\hline
10 & 45,951 & 11 & 30,000 non-citizen voter registration notices, Colorado & No\\
\hline
\end{tabular}
\caption{Top rumors by tweet volume. The top two (which account for over 18.5\% of the data volume), and an additional three of the top ten rumors are focused on Arizona.}
    \label{tab:top_rumors}
\end{table*}

\subsection{Geographic Distribution of Conversation} \label{sec:geography}
We explore the distribution of geographically-focused tweets with two approaches. First, we use the rumor-level coding to propagate state assignments to tweets. Second, we search for references to state names and abbreviations in the tweet text, using regular expressions to find matching substrings. This regular expression search process was also performed on the dataset from \citep{Kennedy2022-qv} regarding the 2020 U.S. presidential election, allowing for comparison. The process used for finding these references is documented in greater detail in \cite{wack_working_2023}. To be clear, the goal of this process is to identify relevant geographic locations mentioned within discussions of the rumor \textit{topic} itself, not the geographic location of the \textit{account} posting about the rumor. Table 
\ref{tab:states} shows the top eight most-commonly-referenced states by approach.

\begin{table}[]
\renewcommand*\arraystretch{1.2}
\begin{tabular}{|l|ll|ll|ll|}
\hline
   & \multicolumn{2}{l|}{2022 Incidents}       & \multicolumn{2}{l|}{2022 References}    & \multicolumn{2}{l|}{2020 References}    \\
Rank       & \multicolumn{1}{l}{State}   & Percentage & \multicolumn{1}{l}{State} & Percentage & \multicolumn{1}{l}{State} & Percentage \\ \hline
1      & \multicolumn{1}{l}{AZ}      & 42.1\%     & \multicolumn{1}{l}{AZ}    & 34.7\%     & \multicolumn{1}{l}{MI}    & 22.0\%     \\ 
2      & \multicolumn{1}{l}{General} & 16.7\%     & \multicolumn{1}{l}{FL}    & 12.9\%     & \multicolumn{1}{l}{PA}    & 21.0\%     \\ 
3      & \multicolumn{1}{l}{FL}      & 9.1\%      & \multicolumn{1}{l}{PA}    & 9.7\%      & \multicolumn{1}{l}{GA}    & 18.4\%     \\ 
4      & \multicolumn{1}{l}{PA}      & 6.8\%      & \multicolumn{1}{l}{MI}    & 5.7\%      & \multicolumn{1}{l}{TX}    & 5.6\%      \\ 
5      & \multicolumn{1}{l}{TX}      & 3.9\%      & \multicolumn{1}{l}{TX}    & 5.3\%      & \multicolumn{1}{l}{AZ}    & 4.7\%      \\ 
6      & \multicolumn{1}{l}{MI}      & 3.0\%      & \multicolumn{1}{l}{IN}    & 4.9\%      & \multicolumn{1}{l}{NV}    & 4.6\%      \\ 
7      & \multicolumn{1}{l}{MO}      & 2.7\%      & \multicolumn{1}{l}{GA}    & 4.1\%      & \multicolumn{1}{l}{CA}    & 3.9\%      \\ 
8      & \multicolumn{1}{l}{WI}      & 2.4\%      & \multicolumn{1}{l}{CO}    & 3.8\%      & \multicolumn{1}{l}{IN}    & 3.4\%      \\ 
Other & \multicolumn{1}{l}{}        & 13.4\%     & \multicolumn{1}{l}{}      & 19.0\%     & \multicolumn{1}{l}{}      & 16.4\%     \\ \hline
\end{tabular}
\caption{The distribution of rumor-related tweets referencing U.S. states, as calculated by both rumor-level state labels and by in-tweet references in 2022, and by in-tweet references in 2020. All states outside of the top eight for a particular year are bucketed into "Other". Note, the percentages for rumor-level labels are relative to all tweets, while for direct references they are only relative to all tweets containing a direct reference to a state.}
\label{tab:states}
\end{table}

\begin{figure}
\centering
\includegraphics[width=.8\linewidth]{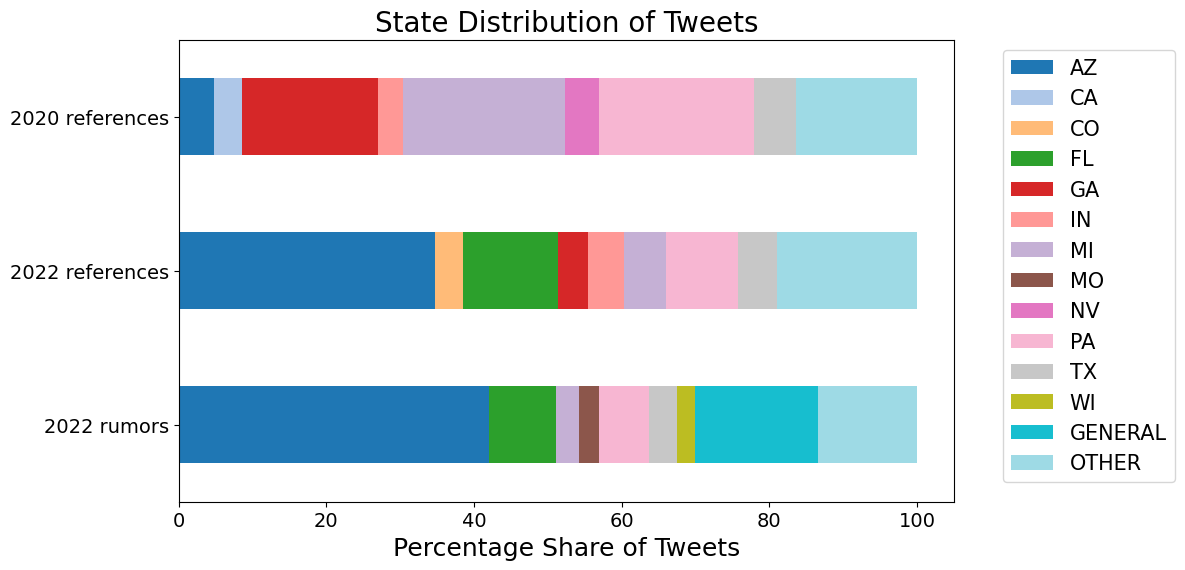}
\caption{Relative distribution of tweets about U.S. states in 2020 and in 2022 by both direct references through substring searches and rumor-level coding. Each row is populated by its eight most frequently referenced states, with the rest bucketed into other. Tweets mentioning multiple states were counted for each state. Note, the percentages for direct reference searches are relative to the total number of tweets referencing a state, not the entire dataset.}
\label{fig:states}
\end{figure}

As shown in Figure \ref{fig:states}, the concentration of narratives on the state of Arizona is far higher in 2022 than it was in 2020, when the state was only ranked fifth and was mentioned in less than five percent of tweets directly referencing a state. Additionally, Arizona in 2022 had a higher concentration than any state did in 2020, where the top states of Michigan, Pennsylvania, and Georgia were far more comparable in terms of rumor volume. We discuss possible explanations for Arizona's prominence in election rumoring in a section below, titled Arizona Rumor Case Studies. 

\subsection{Partisan Distribution of Conversation}
To compare partisan splits within the data, we apply a network clustering method based on audience coengagement \citep{Beers2023-qe}, similar to the approach used in \cite{Kennedy2022-qv}. In brief, we constructed a coengagement projection \citep{Beers2023-qe}\footnote{The coengagement projection results in a network where edges represent retweeting by at least 50 shared audience members between two account nodes, and at least 2 retweets per audience member.} of the full election tweet set (not just those connected to rumors), which identified a subset of highly-retweeted, prominent accounts; we then separate these accounts into clusters using the Louvain method \citep{blondel_fast_2008}. The two large clusters identified align with the political left and political right, which we verify through manual inspection of a sample of the accounts in each cluster. This method provides a set of highly retweeted prominent left-leaning and prominent right-leaning accounts. We then propagate the partisan labels of these prominent account clusters out to the rest of the accounts in the dataset, using retweets as a signal of endorsement. Accounts were marked as likely left-leaning if over 80\% of their retweets of identified prominent accounts were of left-leaning accounts, and right-leaning if over 80\% of their retweets of prominent accounts were of right-leaning accounts. Users with fewer than 80\% of their retweets of prominent accounts linked to a specific partisan cluster were not designated as partisan-aligned. 

This approach enabled us to assign partisanship labels to 388,401 accounts (91\% of all accounts which account for 97\% of all posts in the dataset). We further designate a rumor as partisan-leaning if a majority (over 50\%) of tweets related to that rumor came from users that partisan label. We find that in total 18\% of all tweets and 16\% of all rumors were linked to left-leaning accounts, while 79\% of all posts and 81\% of all 2022 rumors about election processes and results were linked to right-leaning accounts. 

\begin{table}
    \small
    \centering
    \renewcommand*\arraystretch{1.3}
    \begin{tabular}{|c|c|c|c|} \hline 
         Partisanship&  Posts&  Rumors& Accounts\\ \hline 
         Left-Leaning&  327,564 (18\%)&  22 (16\%)& 125,915 (29\%)\\ \hline 
         Right-Leaning&  1,430,244 (79\%)&  109 (81\%) & 262,448 (61\%)\\ \hline 
         Undetermined&  52,833 (3\%)&  4 (3\%)&  39,235 (9\%) \\ \hline
    \end{tabular}
    \caption{The distribution of posts, rumors, and accounts by their assigned partisan affiliation.}
    \label{tab:partisanship}
\end{table}

\begin{figure}
\centering
\includegraphics[width=.9\linewidth]{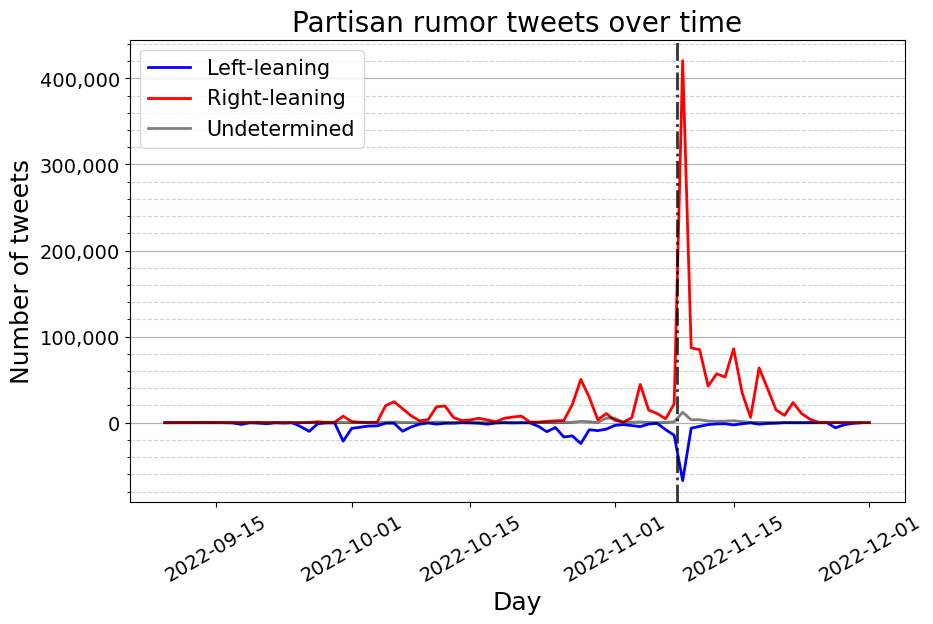}
\caption{A timeline showing the distribution of tweets authored by users in each partisan category. For readability, we represented left-leaning account activity with a downward-facing line, and right-leaning account activity with an upward-facing line. The vertical dashed line represents the start of Election Day.}
\label{fig:partisan}
\end{figure}

\begin{figure}
\centering
\includegraphics[width=.9\linewidth]{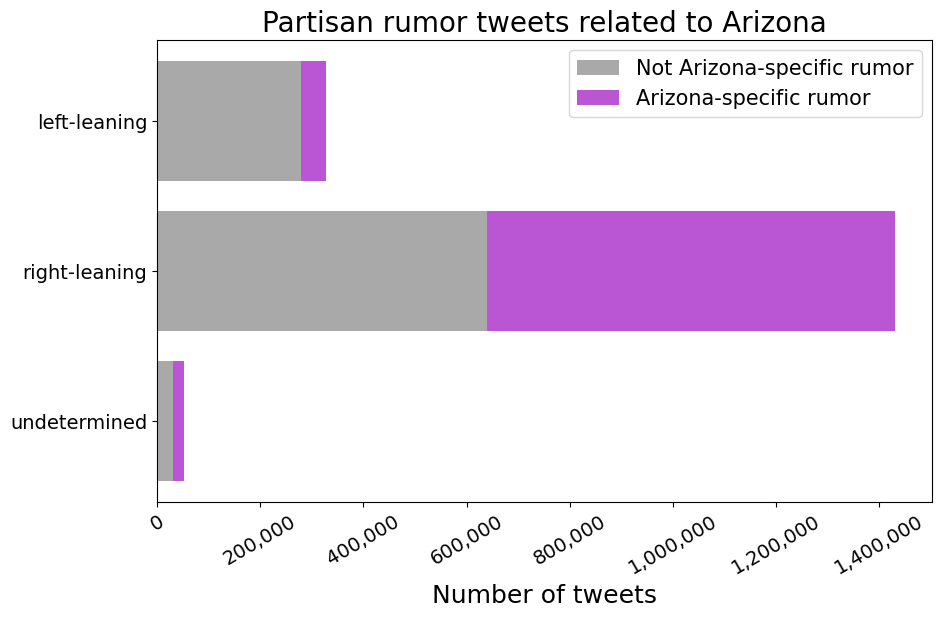}
\caption{A bar graph showing the relative prevalence of tweets in rumors focusing on Arizona in 2022, grouped by the partisanship of users.}
\label{fig:arizona}
\end{figure}

As Figure \ref{fig:partisan} illustrates, the spike in rumoring on Election Day seen in Figure \ref{fig:timeline} is present for both groups, but is particularly evident among right-leaning accounts. By the end of November 2022, most ongoing rumors had dissipated on the platform for both groups. Partisan differences in rumor-engagement are largely similar to those documented in the 2020 election, with right-leaning accounts showing more activity than left-leaning accounts or accounts of undetermined leaning. In this study, we additionally combine the partisan and geographic analyses which reveals another distinct difference between left-leaning and right-leaning groups in 2022. As illustrated in \ref{fig:arizona}, a prominent difference is found in the level of attention given to rumors focused on voting processes and outcomes in the state of Arizona. Examining engagement of each partisan group separately, we find that over half of all posts from right-leaning users in our dataset are related to rumors focused on Arizona (as coded at the rumor level). 
In contrast, only 14.7\% of posts by left-leaning users focused on that state. This chart also partially explains some of the discrepancies between partisan groups' participation in election rumors in 2022. When only looking at tweets which reference general election rumors about the midterms, or are focused on other states, we see somewhat closer partisan splits, though there are still more than twice as many posts from right-leaning accounts than left-leaning accounts.

\subsection{External Links and Media}

In addition to text-based content, Twitter users may link to external websites or embed photos, videos, or GIFs in their posts. Users may be motivated to share links with their audience to inform them of breaking news, to connect with like-minded users or to seek new information \citep{holton2014seeking}. In this section, we analyze the prevalence of posts sharing links to external sources, the diversity and frequency of what domains are linked to, and note key changes from link sharing behavior during the 2020 presidential election. We additionally assess the prevalence of embedded media in rumor-related tweets.

We find that 273,965 (15.13\%) of the tweets within the dataset contain links to external websites. We identify 8,779 unique URLs from 2,057 unique domains (process described in the ``External Links Enrichment" section). Overall, we find that sharing links is highly skewed toward a few of the most popular sites, with the 10 most popular domains accounting for more than half (55.75\%) of the tweets containing URLs. The most popular domains, shown in Table \ref{tab:top domains}, were a mix of news sites and government sites, along with one Republican fundraising platform. 

\begin{table}[h]
\centering
    \renewcommand*\arraystretch{1.2}
    \small
    \begin{tabular}{|l|r|r|c|c|}
        \hline
        Domain & Tweets & Percent & Type & MBFC Bias \\
        \hline
        \texttt{thepostmillennial.com} & 45622 & 14.46 & News & Right \\
        \texttt{thegatewaypundit.com} & 32077 & 10.16 & News & Extreme Right \\
        \texttt{maricopa.gov} & 22721 & 7.20 & Local Gov. &\\
        \texttt{foxnews.com} & 19176 & 6.08 & News & Right \\
        \texttt{secure.winred.com} & 12678 & 4.02 & Fundraising & \\
        \texttt{washoelife.washoecounty.gov} & 9800 & 3.11 & Local Gov. & \\
        \texttt{dailysignal.com} & 9422 & 2.99 & News & Right \\
        \texttt{apnews.com} & 8188 & 2.59 & News & Left Center \\
        \texttt{washingtonpost.com} & 8175 & 2.59 & News & Left Center \\
        \texttt{justice.gov} & 8053 & 2.55 & Federal Gov. & \\
        \hline
    \end{tabular}
    \caption{Most linked domains in rumor-related tweets. The percentage column reports the prevalence of each domain in relation to tweets containing an external link. The bias as listed by Media Bias Fact Check (MBFC)\citep{MBFCheck_2024} are included for sources where that information is available.} 
    \label{tab:top domains}
\end{table}

It is important to note that links to external sites can play a variety of roles in relation to the rumor they are associated with. In some cases, a linked source is in support or promotion of the rumor claims, while in other cases an external source is linked to provide a correction or fact-check. In another scenario linked sources provide a basis of information that the rumor builds upon. Here, we provide several examples of link-sharing behavior that highlight how external links serve different purposes. 

First, is a rumor about the discrepancy between Election Day voter turnout and overall vote totals in Maricopa County, Arizona. The rumor suggests that the fact that Election Day voting had a low proportion of Democrats, but the Democratic candidate for governor (Katie Hobbs) received higher overall vote totals indicates Democratic election malfeasance. In reality, mail-in and early voting trends provide logical explanations as to why Election Day vote splits do not match overall vote splits. The tweet shown below is shared directly from the external source (\texttt{thegatewaypundit.com}). In this case, the linked content is in direct support of the rumor with the article casting the voting statistics as evidence of nefarious behavior. 

\begin{quote}
    \small
    {\fontfamily{qcr}\selectfont @USERNAME-REDACTED: IMPOSSIBLE: Despite Only 17\% Democrat Turnout on Election Day - Katie Hobbs and Democrats Are Winning Over 50\% of Maricopa County Election Day Totals https://t.co/jc7PRrWthy via @gatewaypundit}
\end{quote}

In a second example, we consider one of the three U.S. government (\texttt{.gov}) sites in the list of the ten most-shared domains (Table~\ref{tab:top domains}). In the case of the \texttt{maricopa.gov} domain, the vast majority of links to the site were in relation to two rumors surrounding issues with in-person voting in Maricopa county on Election Day. One of these rumors (further detailed in Case Study 1) includes general speculation about the delays in vote tabulation. The other rumor specifically casts doubt on the policy of mitigating the machine issues by having voters place ballots in a secure drop box (called `box 3' or `door 3') to be counted later. Several highly retweeted tweets by prominent accounts included links to Maricopa's election site that provided information on the location of polling stations. The text of the tweet supports the rumor that the voting procedures cannot be trusted while the link itself provides trustworthy information about polling sites. We include one such example below:

\begin{quote}
\small
{\fontfamily{qcr}\selectfont@kelliwardaz: BIG problems with 
@MaricopaVote. Tabulator ``malfunctions” at at least 6 places. DO NOT PUT YOUR BALLOT IN ``BOX 3” TO BE ``TABULATED DOWNTOWN.” Maricopa will not be turning on the downtown tabulators today. Find your next nearest polling place here: http://Maricopa.vote/ 
}
\end{quote}

Finally, we show an example where factual reporting about a mistake from Colorado's secretary of state office sparked widespread rumors. In this case, the Associated Press (AP) published an article about how a postcard encouraging recipients to register to vote was mistakenly sent to a group of Colorado residents that included a large number of non-citizens in addition to the U.S. citizens that should have received them. The AP shared the article, which included reporting on how the error had occurred and clarified that this did not enable recipients to register unlawfully, on Twitter: 

\begin{quote}
    \small

    {\fontfamily{qcr}\selectfont@AP: Colorado's secretary of state office says it mistakenly sent postcards to about 30,000 noncitizens encouraging them to register to vote, blaming the error on a database glitch related to the state's list of residents with driver's licenses. \\https://t.co/dXWXDhwo83}
\end{quote}

Some users retweeted the AP's post sharing the article, while other users added their own commentary via quote-tweeting the original AP tweet or others who shared the same article. Commentary ranged from criticism of the error, to questioning of circumstances, to conspiracy theorizing, as seen below.  

\begin{quote}
\small
  {\textbf{Comment 1:} \fontfamily{qcr}\selectfont@USERNAME-REDACTED: This oopsie could end up costing these folks their future citizenship. This is appalling and outrageous. Government should take all means possible to inform and stop these noncitizens from wrongly casting votes and jeopardizing their futures.
}
\end{quote}

\begin{quote}
\small
 {\textbf{Comment 2:} \fontfamily{qcr}\selectfont
@USERNAME-REDACTED: Do we actually believe this was a mistake?}
\end{quote}

\begin{quote}
\small
 {\textbf{Comment 3:} \fontfamily{qcr}\selectfont
@USERNAME-REDACTED: Intentional act by a Soros operative? }
\end{quote}

The above three quote tweets reveal a wider pattern of how rumoring interacts with news, where quoted tweets contain factual evidence about election-related events, and commenting tweets integrate that evidence into false, misleading, or unsubstantiated rumors. Taken with the other two examples of shared external links, we see how incorporation of exogenous sources can take a variety of forms in the creation and spread of online rumors. 

\begin{figure}
\centering
\includegraphics[width=.7\linewidth]{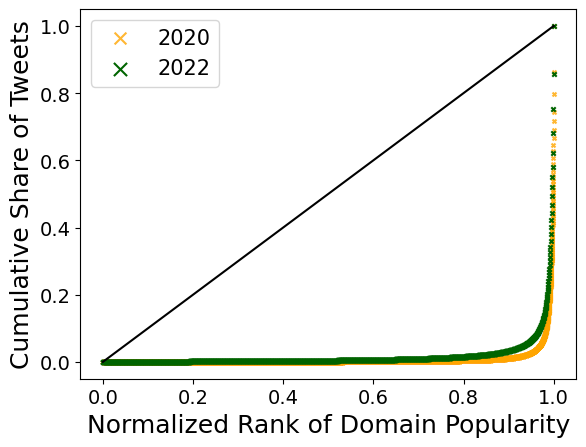}
\caption{Lorenz curve for the popularity of domains shared within rumor or misinformation related tweets.}
\label{fig:domains}
\end{figure}

When compared to the 2020 election, the sharing of external links in rumor-related tweets was far less common. In 2022,  15.14\% of tweets included a link to an external site, compared to 30.33\% of tweets linking to an external site in the 2020 dataset. However, the general pattern of skewed attention to a small number of popular domains held in both cases, as shown in Figure \ref{fig:domains} where the a large number of domains receive a very small cumulative share of tweets and a few popular domains receive the majority of the tweets. Calculating the Gini coefficient as a summary of how attention (via tweets) is split across domains, we find that there was only slightly less domain-level inequality in link-sharing in 2022, with the Gini coefficient falling from 0.983 in 2020 to 0.965 in 2022. 

In Table \ref{tab:domain_change} we show the domains with the most marked difference in popularity between 2020 and 2022. In right-biased news sources, \texttt{breitbart.com} dropped notably in terms of the proportion of links and \texttt{thepostmillenial.com}, which had essentially no references during the 2020 election, was the most popular site shared in 2022 election-related rumors. Interestingly, the official Twitter account for the \texttt{thegatewaypundit.com} was banned in February of 2021 following violations of Twitter's then-active civic integrity policy \citep{twitter_civic_2021}, and remained off the platform for the duration of the 2022 election cycle. However, links to the site were still common; it was the second most shared domain in our dataset despite dropping slightly in relative popularity from 2020. Another notable change is the decrease in links to YouTube, which represented over 6\% of links in 2020 and less than one percent of links in 2022.

\begin{table}[]
\renewcommand{\arraystretch}{1.2}

    \centering
        \begin{tabular}{lrrr}
        \hline
         domain & 2020 & 2022 & Change \\
        \hline
\texttt{thepostmillennial.com} & 0.11\% & 14.46\% & +14.35 \\
\texttt{maricopa.gov} & 0.42\% & 7.20\% & +6.78 \\
\texttt{foxnews.com} & 0.69\% & 6.08\% & +5.38 \\
\texttt{secure.winred.com} & 0.28\% & 4.02\% & +3.73 \\
\texttt{washoelife.washoecounty.gov} & 0.00\% & 3.11\% & +3.11 \\
.. & .. & .. & .. \\
\texttt{inquirer.com} & 2.03\% & 0.02\% & -2.01 \\
\texttt{nationalfile.com} & 2.44\% & 0.01\% & -2.43 \\
\texttt{thegatewaypundit.co} & 13.65\% & 10.16\% & -3.49 \\
\texttt{breitbart.com} & 5.50\% & 0.59\% & -4.91 \\
\texttt{youtube.com} & 6.60\% & 0.71\% & -5.89 \\
        \hline
        \end{tabular}
    \caption{Domains with the largest change in relative popularity between 2020 and 2022. Change is presented as the raw difference in proportion of tweets linking to the domain with positive change signifying an increase in popularity and negative values signifying a decrease.}
    \label{tab:domain_change}
\end{table}

\begin{figure}
\centering
\includegraphics[width=.9\linewidth]{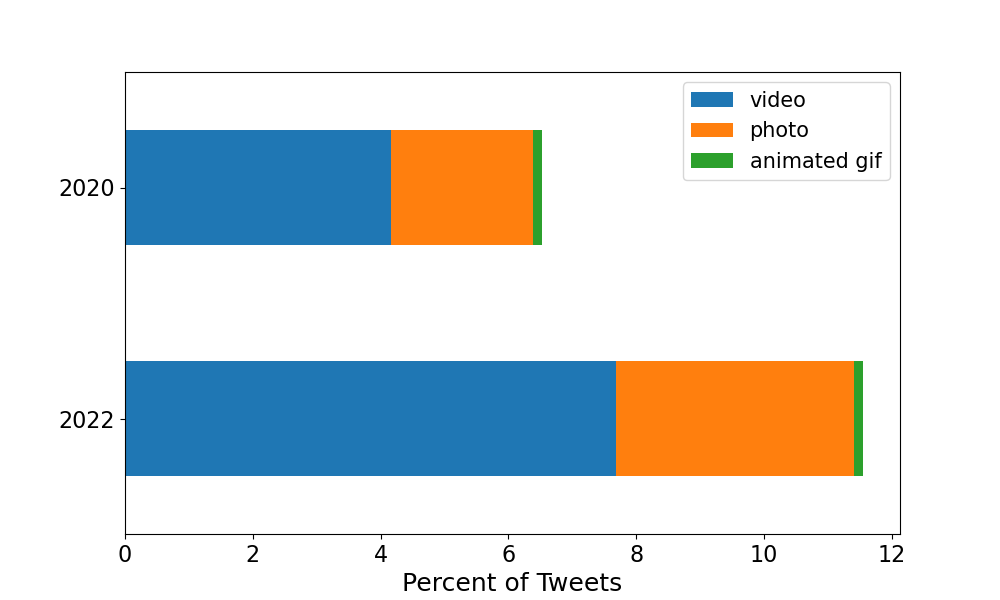}
\caption{Prevalence of embedded media in rumor-related tweets in 2020 and 2022.}
\label{fig:media}
\end{figure}

Twitter enables users to embed photos, videos and GIFs directly in their posts through the use of the `native' media affordance. Rather than linking to an external site that hosts images or videos, users can directly upload media content from their device into their posts. Shown in Figure \ref{fig:media}, we assess the prevalence of these forms of media and find that more rumor-related posts contained `native' media in 2022 than in 2020, increasing from 6.52\% to 11.54\%. This increase in the use of embedded media may help explain the drop in link sharing, as well as the decrease in links to YouTube, though future research designed to understand causal impacts, rather than just the descriptive analysis done here, would be needed to confirm this. 

\subsection{Concentration of Engagement}

One of the key findings of \citep{Kennedy2022-qv} was that original posts from a small number of heavily retweeted and highly active ``repeat spreader" accounts had an outsized contribution to the overall volume of online discourse around false and misleading narratives in 2020. Here, we analyze whether a similar level of retweet concentration was present in 2022. The content produced by relatively few users may receive a disproportionately large amount of engagement through retweets. 

To measure this, we calculated the Gini coefficients for both 2020 and 2022 based on the number of retweets that each user received. We also compute the proportion of total retweets for the top 1\% most-retweeted users, and the top 0.1\% of most-retweeted users, in both 2020 and 2022. These were calculated in aggregate, as well as for each identified partisan cluster;results are shown in Table \ref{tab:influencers}. We observe that, overall, election rumors are highly skewed in their sources toward a relatively small proportion of users. There is no notable change in overall skew (gini coefficient) for any of the groups between 2020 and 2022, and very minor change in the proportion of retweets of the top 1\% most-retweeted users between election cycles. For both left-leaning users and right-leaning users there is a drop in terms of the proportion of retweets of the very select few users who make up the 0.1\% most retweeted population. 

\begin{table}[]
\renewcommand{\arraystretch}{1.2}
\begin{tabular}{|l|ll|ll|ll|}
\hline
      & \multicolumn{2}{c|}{gini} & \multicolumn{2}{c|}{top 1\% users} & \multicolumn{2}{c|}{top 0.1\% users} \\
      
year  & \multicolumn{1}{c}{2020} & \multicolumn{1}{c|}{2022} & \multicolumn{1}{c}{2020} & \multicolumn{1}{c|}{2022} & \multicolumn{1}{c}{2020} & \multicolumn{1}{c|}{2022} \\ \hline
         All & 0.97 & 0.96  & 86\% & 84\% & 58\% & 40\%\\
        Left-leaning & 0.95 & 0.95 & 79\% & 81\% & 41\% & 35\%\\
        Right-leaning & 0.97 & 0.97 & 86\% & 85\% & 58\% & 41\%\\ \hline
\end{tabular}
    \caption{Measures of the level of concentration of retweets of a small number of users among all users, left-leaning users, and right-leaning users. Measurements are the gini coefficient of retweet counts, the proportion of retweets of the top 1\% most retweeted users, and proportion of retweets of the top 0.1\% of users.}
    \label{tab:influencers}
\end{table}

\section{Arizona Rumor Case Studies} \label{sec:az_case_study}

In this section, we feature in-depth mixed-methods analysis of three Arizona-based rumors, a prominent location in our geographic analysis. First, we provide some brief background to help explain why the state of Arizona, and in particular the county of Maricopa, were so salient in our data.

\subsection{Context for Arizona's Prominence in Election Discourse}
In recent elections, Arizona has been considered a ``purple'' or ``swing state'' with fairly similar numbers of Republican and Democrat voters. This means that elections in Arizona are likely to be close, with results potentially impacting the balance of power at the national level. Close elections correlate with higher uncertainty about outcomes, and uncertainty is known to lend itself to rumoring \citep{bordia_psychological_2005}. Additionally, issues with voting, intentional or not, have the potential to impact the results in close elections --- indicating high relevance of the topic for both sides.  

In 2020, Arizona --- and in particular its most populous county, Maricopa County --- became a flashpoint for rumors about election integrity, such as claims about Sharpie pens invalidating votes \citep{leingang_fact_2020} and the Dominion voting systems \citep{center_for_an_informed_public_long_2021}\footnote{Rumors in 2020 about Dominion Voting systems were prominent in Arizona as well as other swing states, including Pennsylvania and Georgia.}. A number of Arizona Republican political figures organized and/or participated in ``Stop the Steal'' protests after the election \citep{shepherd_driven_2020}. Several political operatives and lawyers participated in --- and were indicted for --- a ``fake elector'' scheme that attempted to change the results of the presidential election in Arizona from Biden to Trump \citep{dev_18_2024}. In 2021, an unofficial ``audit'' of Arizona's 2020 general election drew national attention and contributed to sustained distrust in election integrity within the state \citep{clark_cyber_2022}. 

In 2022, gubernatorial candidate Kari Lake promoted theories that elections were rigged by Democrats, using rhetoric similar to Donald Trump (who endorsed her). Lake repeatedly criticized election officials, including her gubernatorial opponent, Katie Hobbs, who was then the Secretary of State of Arizona. 

Going into the 2022 midterm election, this combination of factors likely contributed to widespread distrust, especially among Republicans, in election integrity. Prior to Election Day, false and/or unsubstantiated claims about different elements of the election were already spreading. Then, on Election Day, real problems with voting across the state combined with existing distrust to catalyze dozens of rumors. Some of those rumors --- e.g. that voting machines were not working at many locations --- were true. But others wove emerging news about real issues into unsubstantiated conspiracy theories, e.g. alleging that the problems were an intentional effort to disenfranchise Republican voters.

\begin{figure*}
\centering
\includegraphics[width=.95\linewidth]{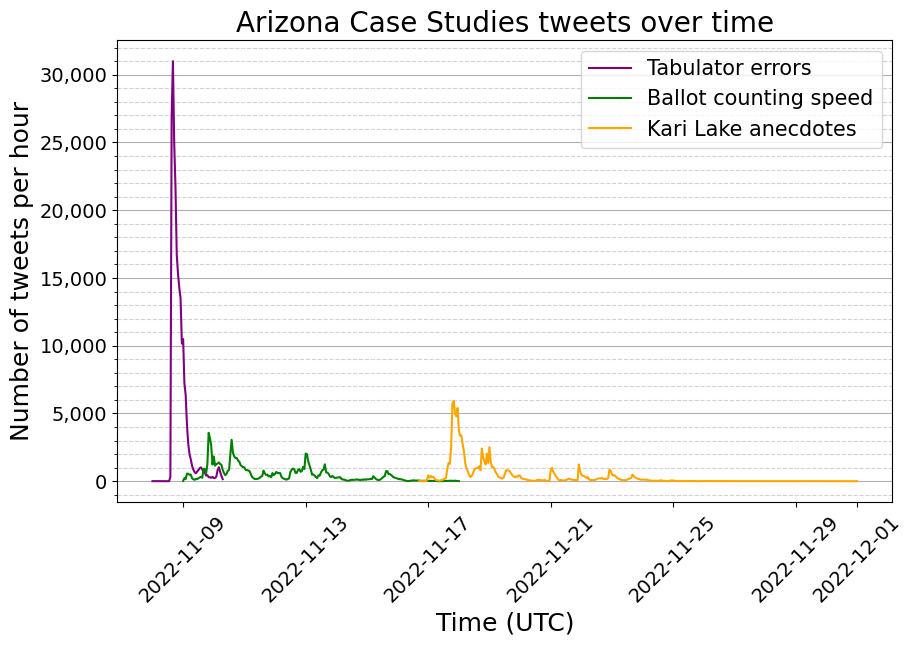}
\caption{A composite timeline of three AZ rumors illustrating their relative timing and volume. As seen, the three rumors show peak volume at different points in time, as well as differences in prominence and longevity.}
\label{fig:az_case_composite}
\end{figure*}

\begin{figure*}
\centering
\includegraphics[width=.95\linewidth]{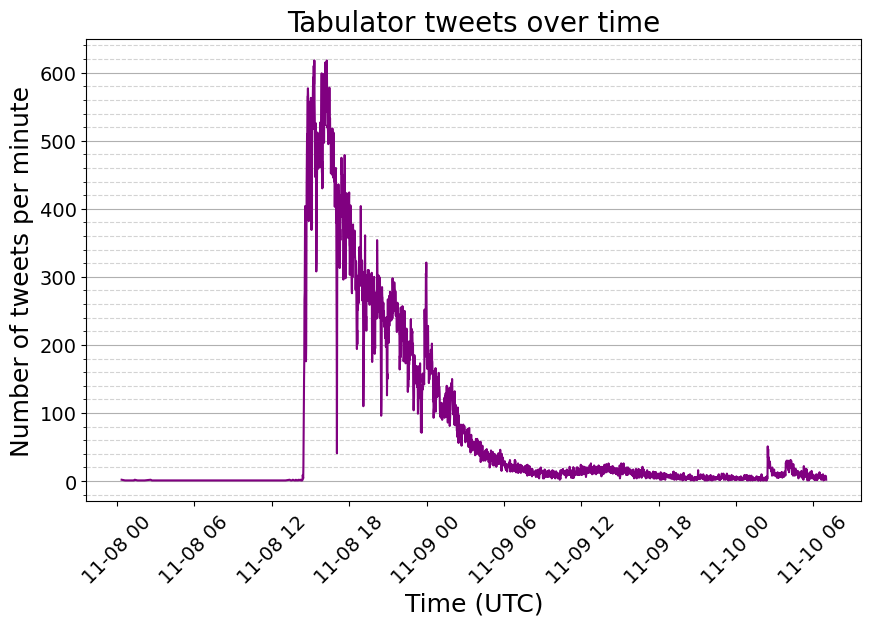}
\caption{A temporal plot showing the number of tweets per minute related to the Arizona tabulator errors rumor. Note, unlike the other case study plots, we use a y-axis of tweets per \textit{minute} here, not per hour, to get a better sense of volume shifts as this rumor had a shorter lifespan than the other cases.}
\label{fig:az_case_1}
\end{figure*}

\begin{figure*}
\centering
\includegraphics[width=.95\linewidth]{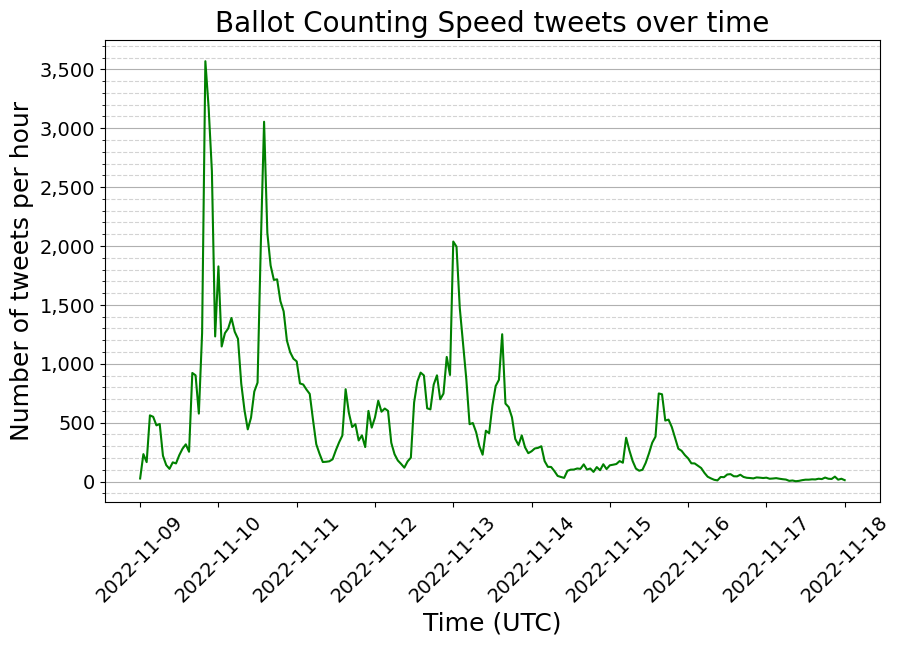}
\caption{A temporal plot showing the number of tweets per hour related to the Arizona ballot counting speed rumor.}
\label{fig:az_case_3}
\end{figure*}

\begin{figure*}
\centering
\includegraphics[width=.95\linewidth]{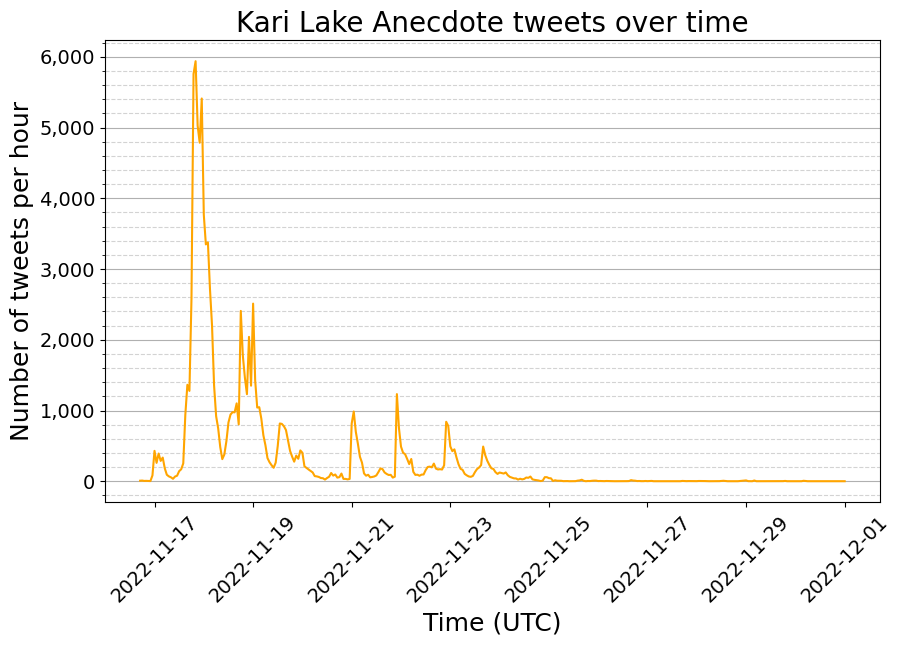}
\caption{A temporal plot showing the number of tweets per hour related to the anecdotes posted by Kari Lake.}
\label{fig:az_case_2}
\end{figure*}

In the following sections, we describe three related rumors that emerged from Maricopa County on and after Election Day. These case studies demonstrate the utility of the described dataset for thorough qualitative work in addition to quantitative methodologies. In Figure \ref{fig:az_case_composite}, we show a timeline of the relative timing and prevalence of all three rumors. Individual timelines for each specific rumor are shown in Figures \ref{fig:az_case_1}, \ref{fig:az_case_3}, and \ref{fig:az_case_2} to provide more detail. 

\subsection{Case Study 1: Tabulators not functioning}
\label{sec:tabulator_case_study}

On Election Day in Arizona, issues with ballots scanners began to occur very early, with reports that scanners were not accepting ballots as early as 6:20 AM local time  \citep{maricopa2022}. Conversation on Twitter began with some limited uncertainty whereby members of the online audience and some Twitter influencers began posting as they described what they knew. For example, one of the earliest highly spread tweets (4,138 retweets) in our dataset came from conservative political operative and COO of Turning Point USA Tyler Bowyer\footnote{Bowyer is one of eleven allegedly fraudulent electors in Arizona who have been indicted for falsely certifying Donald Trump as the winner of the 2020 election in Arizona \citep{barchenger_grand_2024}.}:

\begin{quote}
\small
{\fontfamily{qcr}\selectfont
@tylerbowyer: Long lines in Anthem, Arizona with Poll Workers explaining that the @maricopacounty machines are not working. Do not get out of line!\footnote{In this and subsequent tweets, we have deleted excessive whitespace that made the manuscript unreadable, and added punctuation in brackets if necessary to convey the appropriate meaning.}
}
\end{quote}

Bowyer's tweet embedded a video of a poll worker addressing a long line of voters, describing the issues the polling center was experiencing while also capturing the palpable frustration and skepticism of voters waiting in line. Although Bowyer's tweet primarily shared information about the emerging voting difficulties in Maricopa, members of the online audience commented on the tweet suggesting that the voting issues were evidence of fraud. For example, the following comments were made in response to Bowyer's tweet through the quote tweet affordance:

\begin{quote}
\small
 {\textbf{Comment 1:} \fontfamily{qcr}\selectfont
@USERNAME-REDACTED: I told yall. People looked at me crazy. But I told yall, to have a plan. Also, that if they are going to manipulate, they will do it on election day when all yall are going to vote. The day you have your only chance to vote.
}
\end{quote}

\begin{quote}
\small
 {\textbf{Comment 2:} \fontfamily{qcr}\selectfont
@USERNAME-REDACTED: @RecordersOffice You people shouldn't be allowed to hold elections in your county. Imagine expecting us to believe this isn't intentional. Couldn't get away with the \\sharpies again?
}
\end{quote}

\begin{quote}
\small
  {\textbf{Comment 3:} \fontfamily{qcr}\selectfont@USERNAME-REDACTED: The steal is on where are the Attorneys and the non-biased poll watchers
}
\end{quote}

In each of the above comments, the commenter suggests that the election issues are evidence of a larger, intentional plan. Additionally, \textbf{Comment 1} uses the fact that as rates of mail-in voting have risen in recent elections, Republican voters have voted disproportionately on Election Day to suggest the errors are targeted while \textbf{Comment 2} refers to a previous, false rumor from 2020 in Maricopa County. Specifically, the rumor suggested that Sharpies were being intentionally distributed to invalidate conservative ballots \citep{shepherd_driven_2020}. 

That video and other similar ones were shared in the early hours of rumoring. Initial posts sharing the videos were often fairly neutral in tone, but the videos themselves showed voters who were frustrated and skeptical of the tabulator issues. For example, the most retweeted tweet (with 14,719 retweets in our dataset) within this rumor came from Charlie Kirk, conservative influencer and founder of Turning Point USA, where he embeds the same video Bowyer had shared just minutes earlier:

\begin{quote}
\small
{\fontfamily{qcr}\selectfont@charliekirk11: \emoji{police_light}\emoji{police_light}A poll worker in all-important Maricopa \\County tells Election Day voters the machines are broken. 
}
\end{quote}

Other than calling attention to Maricopa County's political importance, Kirk's tweet is fairly neutral in tone and mainly communicates part of an evolving situation using a video that highlights the frustration and skepticism of voters waiting in line. As the poll worker describes the issues, he states that ``no one is trying to deceive'' and is interrupted by sarcastic commentary from the person recording the event, as well as general groans from those waiting in line. In addition, one voter leaves the line, saying explicitly that they do not trust putting their ballot into the box that is set aside for votes to be tabulated later in the event of machine failure.

Although Kirk's tweet above is relatively neutral at face value, the online audience did not interpret it as such, nor did Kirk and other conservative influencers and political elites shy away from more conspiratorial framing in subsequent discourse. For example, in a tweet quoting Kirk's above tweet, one audience member associates the ongoing tabulator issues with rumors of election fraud in 2020 (known colloquially as the ``Big Lie''):

\begin{quote}
\small
{\fontfamily{qcr}\selectfont@USERNAME-REDACTED: Preparing for the BIG LIE all over again. Explain what's happening here if they don't plan to cheating? Just so happens tabulators aren't working in AZ?
}
\end{quote}

Similarly, in a later tweet, Kirk refers to the tabulator issues as ``manufactured chaos'' and suggests that the resulting lines amount to voter suppression, calling for people to be arrested. This conspiratorial framing was echoed by audiences, influencers, and political elites alike. For example, the seventh most retweeted tweet in the incident (with 6,372 retweets) came from Senator Ted Cruz, who insinuated that then-Secretary of State and Democratic gubernatorial candidate Katie Hobbs was somehow intentionally responsible for the tabulator issues:

\begin{quote}
\small
{\fontfamily{qcr}\selectfont@tedcruz: So, the Dem nominee for governor (who refused to debate her opponent) is the current Secretary of State -- in charge of running this election -- and now... there are problems?\\
\#DemsHateDemocracy
}
\end{quote}

In addition to suspicion around the origin of the errors, online audiences also expressed skepticism as to the reliability of remedies. Maricopa County had backups in place in case of failures like those experienced on Election Day, and in the case of 2022 there was a box on the machines (labeled with the number 3 and referred to as ``Box 3'') where voters were directed to drop their ballots if they were not able to be scanned. However, many members of the online audience expressed skepticism as to whether vote in Box 3 would be accurately recorded. For example, the following tweet insinuates that ballots counted ``downtown'' won't be counted due to corruption:

\begin{quote}
\small
{\fontfamily{qcr}\selectfont @USERNAME-REDACTED Reports of machines not accepting ballots, and that those ballots will be taken "Downtown"... right. Maricopa is as corrupt as they come, and y'all aren't even trying to hide it anymore.
}
\end{quote}

It is important to note that although conspiratorial framing was a major part of the rumoring as it occurred on Twitter, there were many users who simply noted how frustrating the widespread machine failures were but didn't make accusations about fraud. In particular, they highlighted the unpreparedness of the election infrastructure and/or administrators, often describing the failures as the result of incompetence with sentiments similar to the tweet below:

\begin{quote}
\small
{\fontfamily{qcr}\selectfont @USERNAME-REDACTED Monday: Two hour press conference, everything is fine in Maricopa, we know what we're doing. Tuesday: Our machines are broken and we don't know why, but trust our contingency plan. Incompetence is not fraud, but come on with this shit.
}
\end{quote}

Taken together, rumoring on Election Day surrounding tabulator failures was founded on very real issues with election infrastructure. These issues were then used as evidence to support interpretations ranging from suggestions of benign incompetence to claims that the problems were intentional voter fraud perpetrated by Democrats. In support of the conspiratorial interpretation, audiences and influencers highlighted: 1) how more Republicans were voting in person on Election Day than Democrats and were therefore more impacted by machine failures; 2) that the Democratic gubernatorial candidate was acting Secretary of State during her election, insinuating a conflict of interest; 3) that the provided remedy of Box 3 was an attempt to manipulate votes offsite; and 4) that the resulting long lines disenfranchised conservative voters. 

\subsection{Case Study 2: Ballot counting speed}

The second incident we examine surrounds rumoring suggesting that counting ballots slowly is suspicious and allows for the Democratic manipulation of votes. In particular, the rumors center around a comparison between the counting of votes in Florida versus in Arizona. Other states are visible in the data, but the vast majority of tweets compare Florida and Arizona, insinuating that it is suspicious that Florida can finish counting ballots more quickly than Arizona despite having many more ballots to count. For example, conservative influencer Tim Young (@TimRunsHisMouth) posted the following tweet:

\begin{quote}
\small
{\fontfamily{qcr}\selectfont @TimRunsHisMouth: Florida not only had millions more ballots to count than Arizona... but the state was also prepping for a hurricane at the same time AND got the counting done in hours. \\What's Arizona's excuse?}
\end{quote}

The above tweet highlights the difference in both counting speed and number of ballots counted between Arizona and Florida, insinuating that it is suspicious for there to be discrepancies. One of the primary reasons that Arizona counted votes more slowly was that voters changed their behavior regarding mail-in ballots; there were over 290,000 mail in ballots dropped off on Election Day (rather than mailing them in prior) in Arizona in 2022, more than a 70\% increase than the number received in 2020 \citep{fifield_why_2022}. This may have occurred due to increased skepticism of the security of mail-in voting in general, and ballot drop boxes in particular in Arizona, which saw large amounts of rumoring suggesting ``ballot mules'' were using drop boxes for fraud in the lead up to the 2022 general election \citep{prochaska_misinformed_2022}. 

Similar rumors circulated in 2020 as well, and largely converged on interpretations that ballots were intentionally counted slowly so that Democrats could ``find'' ballots to allow them to fraudulently win close elections. The incident in 2022 was similar, with many members of the online audience interpreting the delayed counts as an opportunity for election malfeasance, an example of which is visible below:

\begin{quote}
\small
{\fontfamily{qcr}\selectfont @USERNAME-REDACTED: \#Arizona is just corrupt to the core. \#KariLake has obviously won, yet they are desperately searching for more magical ballots from under tables, just like we saw in 2020. @KariLake is an enormous threat to the establishment and will improve Arizona greatly.Don't let her down} 
\small

\textbf{Quoted tweet:} {\fontfamily{qcr}\selectfont @EndWokeness: 
North Carolina: 98\% counted [,] Population: 10.5 million [.]Wisconsin: 99\% counted  [,] Population: 5.2 million [.] Florida: 99\% counted  [,] Population: 22 million [.] Ohio: 97\% counted  [,] Population: 12 million [.] Arizona: 66\% counted  [,] Population: 1.6 million}
\end{quote}

Above, it is clear that the commenter interpreted the speed of counting as evidence of corruption that allegedly occurred in both 2020 and 2022. Similar to the incidents described above, interpretations ranged from viewing the slow counting as evidence of fraud to suggestions that those claiming fraud were just trying to sow doubt about electoral processes. Between the two were some members of the audience who didn't necessarily view the slow counting as fraud, but instead viewed it as evidence that the election processes in Arizona were a mess and needed to be reevaluated.

\subsection{Case Study 3: Kari Lake anecdotes}

One of the primary reasons that Maricopa County, and Arizona in general, was likely a hotbed for electoral rumoring was because of the endorsement of election denialism by candidates running for office. Most notable of this group was Republican gubernatorial candidate Kari Lake. Although Lake participated in numerous rumors on Twitter to varying degrees, the final incident examined here is focused on a subset of tweets coming from Kari Lake's account (@KariLake) or her campaign's account (@KariLakeWarRoom). 

The primary behavior visible in these tweets was the amplification of individual voters' experiences on Election Day in the form of comments based on accompanying embedded videos from voters. The stories shared often included statements implicitly referring to previous rumors surrounding what has become known as the ``Big Lie'' and connecting them to the events on the ground in 2022. For example, in her most retweeted tweet, Lake amplifies the alleged experience of the son of a voter who ran into trouble voting:

\begin{quote}
\small
{\fontfamily{qcr}\selectfont @KariLake: After registering at ASU, Tiffany's son received a text offering him \$250 to rally for Democrats
In line to vote, he was told to leave \& that his vote would not be counted [.] Inside, he was told he was not registered, given a Sharpie, \& told to drop a provisional ballot in box 3.}
\end{quote}

In this tweet, several themes are visible. First, Lake, pulling claims from the embedded video of a woman describing the experience of her son, claims that Democrats (the primary villains in rumors surrounding the Big Lie) offered money to ``rally for Democrats.'' Second, this was followed by the claim that the long lines catalyzed by the tabulator failures described above resulted in poll workers asking voters to leave, allegedly because he, and others, would be unable to vote by the deadline of 7:00 PM. Lastly, Lake highlights that poll workers tried to give the son a Sharpie to fill out his ballot, referencing the previous rumor from 2020.

In addition to the content, the style of the above tweet and most of Lake's tweets in this incident is important --- Lake makes few direct claims herself, instead relying on testimony from voters to interpret emergent events within a frame of voter fraud established in 2020 and amplified by Lake throughout her campaign in 2022. This process is described in more detail in other works \citep{Starbird2023-ht,prochaska_mobilizing_2023}, but at its core consists of establishing an expectation within conservative voters that Democrats will perpetrate fraud, which then causes audiences to organically organize and interpret otherwise ambiguous information as ``evidence'' of fraud instead of something more mundane such as an error in election administration. This ``evidence'' is amplified online, providing ``proof'' that is strategically amplified by political elites and influencers to support the ongoing conversations surrounding alleged voter fraud in U.S. elections.

Seen from this perspective, the rumors spread are the product of an informal collaboration between political elites, influencers, and members of the public, all of whom play integral roles in the production and spread of rumors online. Although many members of the public participate in the production and dissemination of these rumors, the interpretations of Lake's tweets are not uniform. Within this incident, interpretations of Lake's tweets ranged from explicitly viewing the shared anecdotes as evidence of fraud to suggestions that the videos are evidence of the negative consequences of misleading rhetoric, visible in the example below:

\begin{quote}
\small
{\fontfamily{qcr}\selectfont @USERNAME-REDACTED: Here's Larry's story, fresh from Kari Lake. Larry's story is genuinely sad. So spun up on Big Lie bullshit conspiracy theories he apparently refused to submit to put his ballot in the ballot box because he thought it would be thrown away.} 

\small
\textbf{Quoted tweet:} {\fontfamily{qcr}\selectfont @KariLake: It took Larry an hour \& a half to get into his polling location  [.] Inside, Larry's ballot was repeatedly rejected by the tabulator. He was asked to put it in box \\three so it could be counted downtown. He refused. Because he felt it meant they would throw his vote in the trash.} 
\end{quote}

In the above tweet, the poster highlights how election skepticism on the part of voters caused them to doubt the remedy in place in Maricopa County for tabulator issues, namely the use of ``Box 3'' for ballots to be counted at a separate location in the event of machine issues. Although there were a noticeable number of responses countering interpretations of the anecdotes, those interpreting them as evidence of fraud or disenfranchisement were more prominent. Within these responses, users often utilized other related rumors to support their interpretations, including suggestions that Katie Hobbs is too biased to be relied on as Secretary of State, claims that noncitizen voters were voting illegally, that errors on Election Day were intentional, and that swing states are the primary targets for Democrat-led voter fraud.

\subsection{Case study summary}
In the above three cases, we have illustrated a selection of prominent rumors which spread during the 2022 midterms about the election in Arizona. These showed a variety of communicative and discursive practices being deployed to facilitate the spread of these rumors, including differences in tone between influential amplifiers and their audience (Case 1) and the elevation and generalization of anecdotes (Case 3). These cases illustrate the need for further qualitative work on election rumor discourses in future research. 

\section{Limitations}
This dataset attempts to comprehensively cover English-language tweets about false, misleading, and/or unsubstantiated rumors about the election process of the 2022 U.S. midterm elections. However, this scoping means that there are aspects of closely-related questions this dataset should not be used to address. Notably, it does not include non-English language rumors, nor rumors about candidates that are not tied to election processes. Discussions of rumors on platforms other than Twitter are not captured, as data collection access prevents coverage of other platforms with the same level of comprehensiveness. A further form of data missing is rumors which spread primarily through images with no (or minimal) in-text keywords. Additionally, this dataset, as it is focused on discussions of false and misleading rumors specifically, should not be used for more general analyses of broad political discussions during this time, as large parts of political discourses are unaffiliated with the kinds of rumors examined in this paper.

\section{Conclusion}

In this paper, we have documented the collection, curation, and preliminary descriptive analysis of a dataset of 1.81 million election administration-related rumors on Twitter posted around the 2022 U.S. midterms. This data contributes to research into online rumoring around elections by looking into a comparatively under-studied midterm election process rather than presidential elections. After providing detailed documentation of the process for ensuring the reliability of the dataset, we conducted five descriptive analyses of this data, exploring overall tweet frequency, geographic and partisan distributions, the prevalence of external link-sharing, and the highly concentrated nature of attention to a small number of users. We follow this with mixed-methods case studies of three prominent rumors about election issues in Arizona, highlighting how this dataset could be used for both qualitative and mixed-methods future research in addition to the quantitative research styles contained in this paper. 

Our findings collectively demonstrate several interesting features of the dynamics of rumoring during the midterm elections, and how these have changed since 2020 \citep{Kennedy2022-qv}. We do not make causal claims about these shifts, but observe several changes including much higher concentration on one particular state and lower prevalence of content linking to external domains. At the same time, we observe a similar partisan asymmetry to what \citep{Kennedy2022-qv} found in 2020, particularly concentrated in Arizona, and similar levels of retweet concentration.

The spread of rumors online has been an area of ongoing interest for research and for society at large, and understanding the spread of rumors in the context of democratic elections is especially important. This dataset can help support research in this area by providing a wide-ranging overview of rumors that spread within this U.S. election context. 

This dataset could inform research which seeks to understand more granular levels of rumor spread. For example, research into rumor spread in Arizona during the midterms could take the dataset as a starting point for areas to further investigate. Other studies of Twitter content in this period could also use this dataset to understand if and how prevalent these rumors are within their own observations or data. For example, algorithmic audits of Twitter conducted during the election period, such as \citep{duskin_echo_2024}, could use this dataset to estimate the frequency that observed election rumors are shown to Twitter accounts. 

Our dataset of 1.81 million posts related to discussions of false, misleading, or unsubstantiated rumors surrounding the 2022 U.S. midterm elections on the Twitter/X platform provides a thoroughly scoped and curated view of these kinds of discussions on the platform at the time. This will provide utility to researchers seeking to conduct further research on these topics, as well as increase our understanding of election rumor discussions from our empirical findings. 

\subsection*{Ethical statement.}
This data was determined by the Human Subject Division at the University of Washington not to involve human subjects, as defined by federal and state regulations and therefore, did not require review and approval by the IRB. In accordance with Twitter's Terms of Service at the time of collection, the only data that we released which came from the platform are user IDs and tweet IDs - this does not include tweet text, usernames, media, or other fields - and in fact we only release pseudonymized versions of the user IDs. Since the removal of the free research API by Twitter (now X) has made rehydrating potentially more difficult for researchers, we will make the hydrated data available to researchers upon reasonable request, similar to prior datasets such as \citep{Aiyappa2023-rp}. In this paper, analysis of inferred political leaning was conducted at an aggregated community level, which is in line with prior work (see for examples, \citep{Abilov2021-ki, Beers2023-qe,Sharma2022-zx}). 

Similarly to \citep{Giorgi2022-eo}, we consider the ethical questions in Datasheets for Datasets \citep{Gebru2021-tu}. While we did not get users' consent to collect these data, this is consistent with substantial amounts of prior work on public Twitter posts, and by only providing user ID and tweet ID numbers, users who wish to have their content anonymized can delete their data from the platform and this would not be rehydratable. This is similar to other published datasets, including those related to similarly sensitive political topics, such as \citep{Abilov2021-ki, Aiyappa2023-rp}.

\section*{Acknowledgments}
Funding for this work has come from the University of Washington's Center for an Informed Public, the John S. and James L. Knight Foundation (G-2019-58788), Craig Newmark Philanthropies, the William and Flora Hewlett Foundation, the Election Trust Initiative, the National Science Foundation (grant  \#1749815 and grant \#2120496) and NSF Graduate Research Fellowships under Grant No DGE-2140004, for both Joseph S. Schafer and Kayla Duskin. Any opinions, findings, conclusions, or recommendations expressed in this material are those of the authors and do not necessarily reflect the views of the National Science Foundation or other funders. We would also like to acknowledge Alex Loddengaard for their infrastructural support for this project, and Kristen Engel and Ben Yamron for feedback on sections of the writing of this paper.

\bibliographystyle{ACM-Reference-Format}
\bibliography{main}

\appendix
\pagebreak
\section*{Appendix}

\renewcommand{\thetable}{A\arabic{table}}
\setcounter{table}{0} 
\begin{table}[h!]
\renewcommand{\arraystretch}{1.2}
    \centering
    \small
    \begin{tabular}{p{15cm}} \hline 
    BMD, BMDs, EVM, EVMs, HandMarkedPaperBallots, USPS, absentee, adjudication, 
    arizona, audit, ballot, ballots, bmd, bmds, chain of custody, chicago, cisa, cochise, code, 
    codes, color revolution, \textit{conservative},  \textit{conservatives}, decertification, decertify, \textit{dem}, \textit{democrat}, \textit{democrats}, \textit{dems}, desantis, detroit, \textit{DNC}, dominion, drop box, drop boxes, dropbox, dropboxes, election, election2022, electioneering, electionfraud, elections, elections2022, electors, electors, epoll, es\&s, forensic, fortalice, fraud, fraudulent, fulton, \textit{GOP}, \textit{GOPers}, halderman, hand count, hand-count, hand-counted, hand-marked, handcount, handcounted, handcounts, handmarked, imagecast, imagecastx, integrity, intimidation, lancaster, \textit{liberal}, \textit{liberals}, machine, machines, mail, mail in, maricopa, michigan, midterm, midterms,  midterms2022, mule, nomachines, noncitizen, onenightcount, overvote, paperballots, pennsylvania, philadelphia, pima, pinal, poll, pollbook, pollbooks, polling, polls, pollwatcher, pollwatchers, pollworker, post office, postal, postoffice, precinct, racine, raffensperger, recount, \textit{republican}, \textit{republicans}, results, rigged, rigged voterfraud, risk-limiting audit,  \textit{RNC}, rolls, smartmatic, subversion, suppression, tabulator, tabulators, tallies, tamper,  tampered, tampering, touchscreen, touchscreens, undervote, vote, votebymail, voted, voter,  voterfraud, voters, votersuppression, votes, votesuppression, voting, vulnerabilities, vulnerability,  yuma \\ \hline
    \end{tabular}
    \caption{The set of keywords used to create the initial election tweet pool. Terms listed in italics were removed from the keyword list on 11/7/2022. The term `desantis' was added on 11/15/2022.}
    \label{tab:keywords}
\end{table}

\begin{table}[h!]
\renewcommand{\arraystretch}{1.2}
    \centering
    \small
    \begin{tabular}{p{4cm}p{10cm}}
           Code& Description \\\hline
           Insufficient Coverage&  Coverage could not be found on this specific event or class of events \\
           Largely Substantiated&  Coverage exists, and confirms the major elements of online narrative, including impact and cause/motive. \\
           \textbf{Unsubstantiated} &  \textbf{Coverage exists, but neither confirms or debunks major elements of online narrative (if false/misleading elements, choose false/misleading)}\\
           \textbf{False/Misleading}&  \textbf{Coverage exists and highlights false, misleading, or unsubstantiated elements of online narrative.}\\
           No Central Claim&  The incident focuses on a piece of media that advances many claims (such as a longer video or podcast) and no particular claim is central enough to rate.\\
    \end{tabular}
    \caption{Coding scheme for identifying the status of each rumor based on authoritative sources. The two in-scope categories, which were kept for analysis in this paper, are marked with bolded text.}
    \label{tab:coverage_codes}
\end{table}

\subsection*{Note on Source Table Data}

Two of our collection methods for posts were added to improve comprehensiveness --- collections from back-filled tweets, and collections based on key location terms which we anticipated would be the site of significant election rumoring. However, including these slightly changes the composition of the dataset in skewing ways. Approximately 5.6\% of our dataset came from the locations collector, while  approximately 0.08\% came from the back-filling collector. In the analyses above, we chose to use the more comprehensive collection and include all tweets. We ran the geographic analysis without the tweets from the locations collector, and found no distinguishable changes in results. For other researchers focusing on different questions, excluding these tweets may be appropriate - we provide the data on tweets collected solely through these means in the source table.

\end{document}